\documentclass[aps,prl, groupedaddress, twocolumn, superscriptaddress]{revtex4-1}
\usepackage{enumitem}
\usepackage{color}
\usepackage{bpchem}
\usepackage{graphicx}
\usepackage{dcolumn}
\usepackage{bm}
\usepackage{amsmath,amsthm,amssymb}
\usepackage{epstopdf}
\usepackage{float}
\usepackage{color}
\usepackage{color}
\usepackage{hyperref}

\begin{document}

\title{Magnetic Fluctuations, Precursor Phenomena and \\ Phase Transition in MnSi under Magnetic Field}

\author{C. Pappas}
\email{c.pappas@tudelft.nl}
\affiliation{Delft University of Technology, Mekelweg 15, 2629 JB Delft, Netherlands}

\author{L.J. Bannenberg}
\affiliation{Delft University of Technology, Mekelweg 15, 2629 JB Delft, Netherlands}

\author{E. Leli\`evre-Berna}  
\affiliation{Institut Laue-Langevin, 71 Avenue des Martyrs, 38000 Grenoble, France}

\author{F. Qian}
\affiliation{Delft University of Technology, Mekelweg 15, 2629 JB Delft, Netherlands}

\author{C. D. Dewhurst}  
\affiliation{Institut Laue-Langevin, 71 Avenue des Martyrs, CS 20156 Grenoble, France}

\author{R. M. Dalgliesh}
\affiliation{STFC, ISIS, Rutherford Appleton Laboratory, Didcot OX11 0QX, United Kingdom}

\author{D. L. Schlagel}
\affiliation{Ames Laboratory, Iowa State University, Ames, IA 50011, USA}

\author{T. A. Lograsso}
\affiliation{Ames Laboratory, Iowa State University, Ames, IA 50011, USA}

\author{P. Falus}
\affiliation{Institut Laue-Langevin, 71 Avenue des Martyrs, CS 20156 Grenoble, France}

\begin{abstract}
The reference chiral helimagnet MnSi is the first system where skyrmion lattice correlations have been reported. At zero magnetic field the transition at $T_C$ to the helimagnetic state is of first order. Above $T_C$, in a region dominated by precursor phenomena, neutron scattering shows the build up of strong chiral fluctuating correlations over the surface of a sphere with radius $2\pi/\ell$, where $\ell$ is the pitch of the helix. It has been suggested that these fluctuating correlations  drive the helical transition to first order following   a scenario proposed by Brazovskii for liquid crystals. We present a comprehensive neutron scattering study under magnetic fields, which provides evidence that this is not the case. The sharp first order transition persists for magnetic fields up to 0.4~T whereas the fluctuating correlations weaken and start to concentrate along the field direction already above 0.2~T.  Our results thus disconnect the first order nature of the transition  from the precursor fluctuating correlations. They also show no indication for a tricritical point, where the first order transition crosses over to second order with increasing magnetic field. In this light, the nature of the  first order helical  transition and the precursor phenomena above $T_C$, both of general relevance to chiral magnetism, remain an open question.

\end{abstract}


\date{\today ~version~4.0 }

\maketitle
In cubic chiral helimagnets as MnSi, chiral skyrmions are stabilized under magnetic field and form a lattice in the so-called $A$-Phase, a small pocket in the temperature-magnetic field phase diagram just below the critical temperature $T_C$ \cite{muhlbauer2009}. These skyrmions are topologically protected vortex-like spin textures that have been observed both in reciprocal space by neutron scattering and in real space by Lorentz Transmission microscopy \cite{bogdanov1989, rossler2006, muhlbauer2009, munzer2010,yu2010, FeGe_Lorenz_TEM, nagaosa2013}. 

The helical transition at zero magnetic field is of first order \cite{bak1980, stishov2007, pappas2009, pappas2011, janoschek2013} and preceded by strong chiral fluctuating correlations. These build up just above $T_C$ leading to intense diffuse scattering of neutrons that  spreads homogeneously on the surface of a sphere with radius $\tau=2\pi/\ell$, with $\ell$ the pitch of the helix \cite{grigoriev2005}. This peculiar scattering occurs  in a region dominated by precursor phenomena that have been probed with a wide range of experimental techniques \cite{stishov2007,bauer2013,petrova2015,stishov2016,pappas2009,pappas2011}. Similar precursor phenomena have  also been found in FeGe \cite{wilhelm2011,cevey2013} and the multiferroic Cu$_2$OSeO$_3$ \cite{adams2012, sidorov2014, zivkovic2014} and are thus of general relevance to chiral magnetism. 

It has been suggested that the precursor fluctuating correlations drive the helical transition to first order \cite{janoschek2013} according to a scenario originally proposed by Brazovskii for liquid crystals \cite{brazovskii1975}. This approach provides a good description of the temperature dependence of the correlation length and the susceptibility in MnSi \cite{janoschek2013}. However, it does not satisfactorily describe all observations \cite{stishov2016} and is less conclusive for other chiral magnets such as Cu$_2$OSeO$_3$ \cite{zivkovic2014} and Fe$_{0.7}$Co$_{0.3}$Si \cite{bannenberg2017helimagnetic}. Thus the origin of the first order phase transition in helimagnets and its  connection to the fluctuating correlations and  precursor phenomena above $T_C$ is still an open question, that will be addressed in this letter for the case of MnSi. For this purpose we combined Small Angle Neutron Scattering (SANS) and Neutron Spin Echo (NSE) spectroscopy to monitor the influence of a magnetic field on both the phase transition and the fluctuating correlations. The experimental results show that the fluctuations are no longer isotropic  for $B\gtrsim 0.2$~T whereas the sharp first order transition persists up to at least $B\sim $ 0.4~T. It  appears that the first order helical transition and the precursor phenomena are not interrelated as expected by the Brazovskii approach. Consequently,   the nature of the  first order helical  transition and of the precursor phenomena above $T_C$, both of general relevance to chiral magnetism, remain an open question.


The measurements were performed on an almost cubic single crystal of MnSi with dimensions $\sim$5x5x5~mm$^3$ grown from a stoichiometric melt as described in  \cite{stishov2007}. The crystal was aligned with the [1$\bar{1}$0] crystallographic axis vertical. The SANS experiments were performed at the time-of-flight instrument LARMOR located at the neutron spallation source ISIS. The neutron beam was unpolarized and contained neutrons with wavelengths of 0.09 $\, \leq \lambda  \leq \,$1.25~nm. The measurements were performed with the magnetic field applied both along and perpendicular to the incoming neutron beam, designated by the wavevector $\vec{k}_i$, and thus provide a full picture of the magnetic correlations.  All SANS patterns were normalized to standard monitor counts and background corrected using a high temperature reference measurement at 40~K. 

\begin{figure*}
\begin{center}
\includegraphics[width= 1\textwidth]{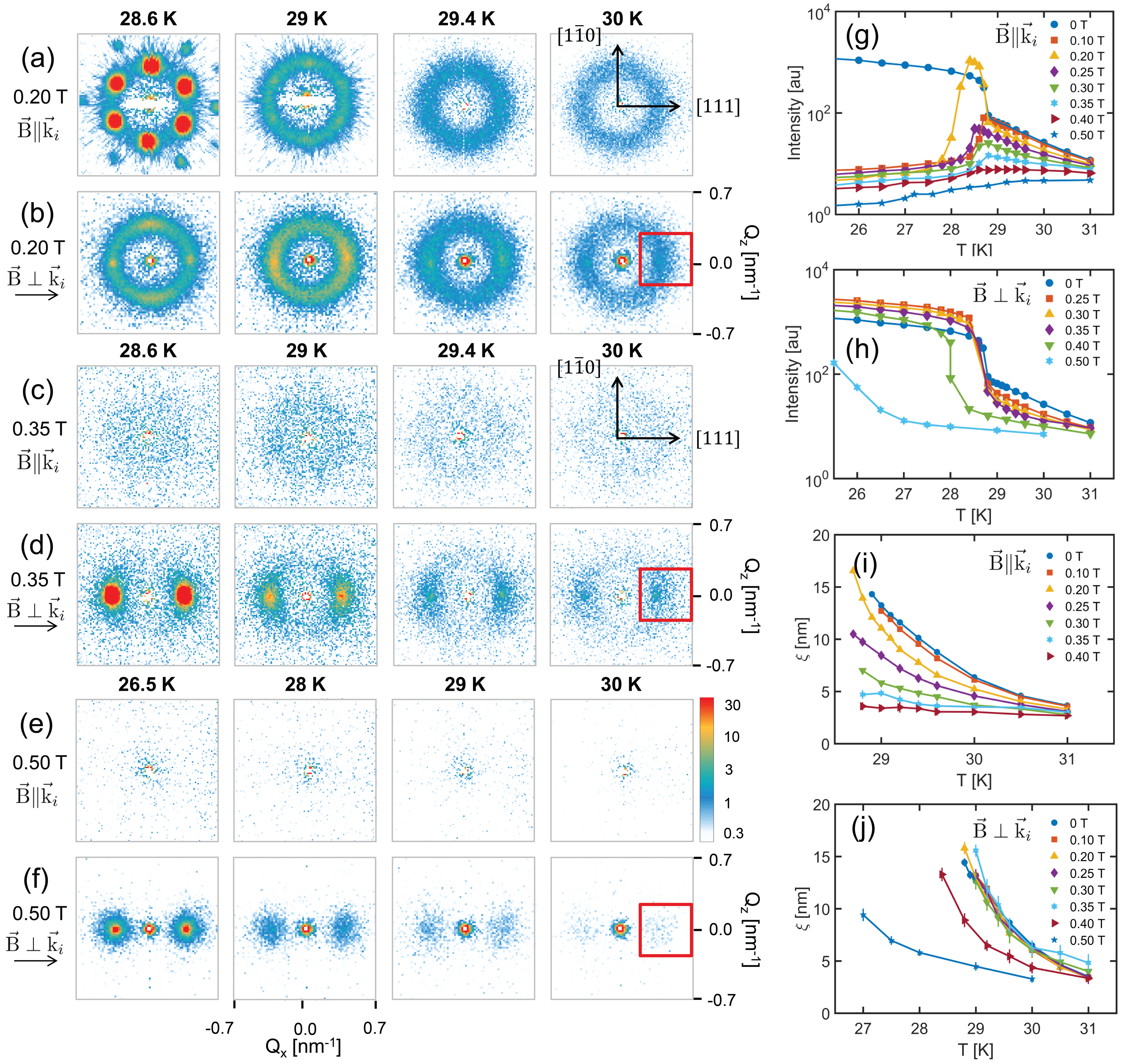}
\caption{SANS scattering patterns obtained for a magnetic field of 0.20~T (a)-(b), 0.35~T (c)-(d) and 0.50~T (e)-(f), where the field was applied parallel to the neutron beam ($\vec{B} || \vec{k}_i$) and perpendicular to it ($\vec{B} \perp \vec{k}_i$). The red squares illustrate the angular acceptance of the NSE experiments. The temperature dependence of the total scattered intensity obtained by summing the entire detector is displayed in Panel (g) for several fields along the neutron beam and in Panel (h) for fields perpendicular to it. Panel (i) and (j) show the temperature dependence of the correlation length for magnetic fields along and perpendicular to the neutron beam, respectively.}
\label{SANS}
\end{center}
\end{figure*}

The NSE experiments were performed at the spectrometer IN15 of the Institute Laue-Langevin, which has a highly polarized incident neutron beam and polarization analysis capabilities but a restricted angular acceptance, as illustrated by the red squares in Fig. \ref{SANS}. The measurements were performed for $\lambda = 0.9$ and 0.8~nm with $\Delta \lambda / \lambda \sim$ 15\%. The magnetic field was applied  perpendicular to $\vec{k}_i$ using a horizontal field cryomagnet. The NSE spectra were recorded in the ferromagnetic NSE configuration \cite{Farago:1986vl, Pappas:2008fb} and were averaged over the entire detector since no significant Q-dependence was found.


SANS measures the static structure factor $S(Q)\equiv S(Q, t=0)$, with $Q$  the momentum transfer, which is the spatial Fourier transform of the correlations. For fluctuating correlations decaying exponentially with distance $d$, typically $\sim \exp(-d/\xi)/d$ with $\xi$ the characteristic correlation length, $S(Q)$ assumes the general Ornstein-Zernike form:
\begin{equation}
S(Q)={C}/\left(\left(Q-\tau\right)^2+\xi^{-2}\right).
\label{eq:OZ}
\end{equation}
With NSE spectroscopy we directly determined the normalized intermediate scattering function $I(Q,t) = S(Q,t)/S(Q)$, thus the decay in time of the correlations. The combination of these two neutron scattering techniques  provided a complete characterization of the magnetic correlations both in space and in time.  

Fig. \ref{SANS}(a) depicts typical SANS patterns for $B = 0.20$~T with $\vec{B} || \vec{k}_i$, a configuration that is sensitive to helical modulations perpendicular to the field. Below $T_C$, the six-fold symmetry scattering pattern shows the existence of the skyrmion lattice \cite{muhlbauer2009} on top of a weak ring of diffuse scattering. The ring of diffuse scattering, similar to the one seen at zero field \cite{grigoriev2007,janoschek2013}, is clearly present above $T_C$ $\sim$ 28.7 K. This feature broadens and decreases in intensity with increasing temperature. 

The patterns in the complementary configuration of $\vec{B} \perp \vec{k}_i$, that is sensitive to helical modulations along the field, are displayed in Fig. \ref{SANS}(b). They show below $T_C$ the ring of scattering superimposed on two peaks oriented along the direction of the field,  which are the signature of the conical phase. The diffuse scattering persists above $T_C$ but, unlike at zero field, it is not homogenous but concentrates along the magnetic field direction.  This effect is accentuated at higher fields shown in Fig. \ref{SANS}(d) and (f) for $B$ = 0.35~T and 0.5~T respectively.  In the complementary configuration  $\vec{B} || \vec{k}_i$, the ring is hardly visible for $B$ = 0.35~T (Fig. \ref{SANS}(c)) and is almost completely suppressed for $B$ = 0.5~T (Fig. \ref{SANS}(e)). 

 
Fig. \ref{SANS}(g) and (h) depict the total scattered intensity  plotted as a function of temperature for $\vec{B} || \vec{k}_i$ and $\vec{B} \perp \vec{k}_i$  respectively.  Both figures show that at zero magnetic field the onset of the helical Bragg peaks at $T_C$ leads to a sharp jump in intensity by more than an order of magnitude within $\sim$0.2~K. Under magnetic field this intensity jump persists but only for $\vec{B} \perp \vec{k}_i$ and  $B$ $\lesssim$ 0.4~T. At  $B$ = 0.5~T this  intensity increase occurs  gradually over more than 1~K. In the complementary configuration, for  $\vec{B} || \vec{k}_i$, the intensity  jump  at $T_C$ becomes a cusp with the exception of $B$ = 0.2~T, where a maximum marks the skyrmion lattice phase. For magnetic fields exceeding  0.2~T, the cusp becomes less prominent and disappears for $B$ = 0.5~T where the magnetic field suppresses most of the scattering.

The effect of the magnetic field on the fluctuations, seen  in the anisotropy of the diffuse scattering discussed above, is also reflected on the deduced correlation lengths, obtained by fitting   the radial averaged intensities, given in the supplement, to Eq. \ref{eq:OZ} convoluted with the instrumental resolution. These are  given in  Fig. \ref{SANS}(i) and (j) for $\vec{B} || \vec{k}_i$  and $\vec{B} \perp \vec{k}_i$ respectively.  For B $<$ 0.2~T, $\xi$ is independent of both the magnetic field and its orientation, and is in good agreement with the values reported in the literature for zero field \cite{Maleyev_MnSi, pappas2009, pappas2011, janoschek2013}:  $\xi$ increases monotonously from $\sim$3 nm at $T_C$+2~K up to approximately 16~nm, i.e. the pitch of the helix  at $T_C$ (see supl. Fig. S2(c)). For B $>$ 0.2~T,  $\xi$ is no longer isotropic and depends on the magnetic field orientation. For $\vec{B} || \vec{k}_i$,  $\xi$ decreases with increasing magnetic field and at 0.4~T it stays practically constant between 29 and 32~K and does not exceed 4~nm. In the complementary configuration, for $\vec{B} \perp \vec{k}_i$, the zero field behavior remains unchanged up to 0.4~T, despite a slight shift of $T_C$ for 0.4~T. At 0.5~T the increase of  $\xi$ persists but stretches over a wide temperature range.



\begin{figure*}
\begin{center}
\includegraphics[width= 0.8 \textwidth]{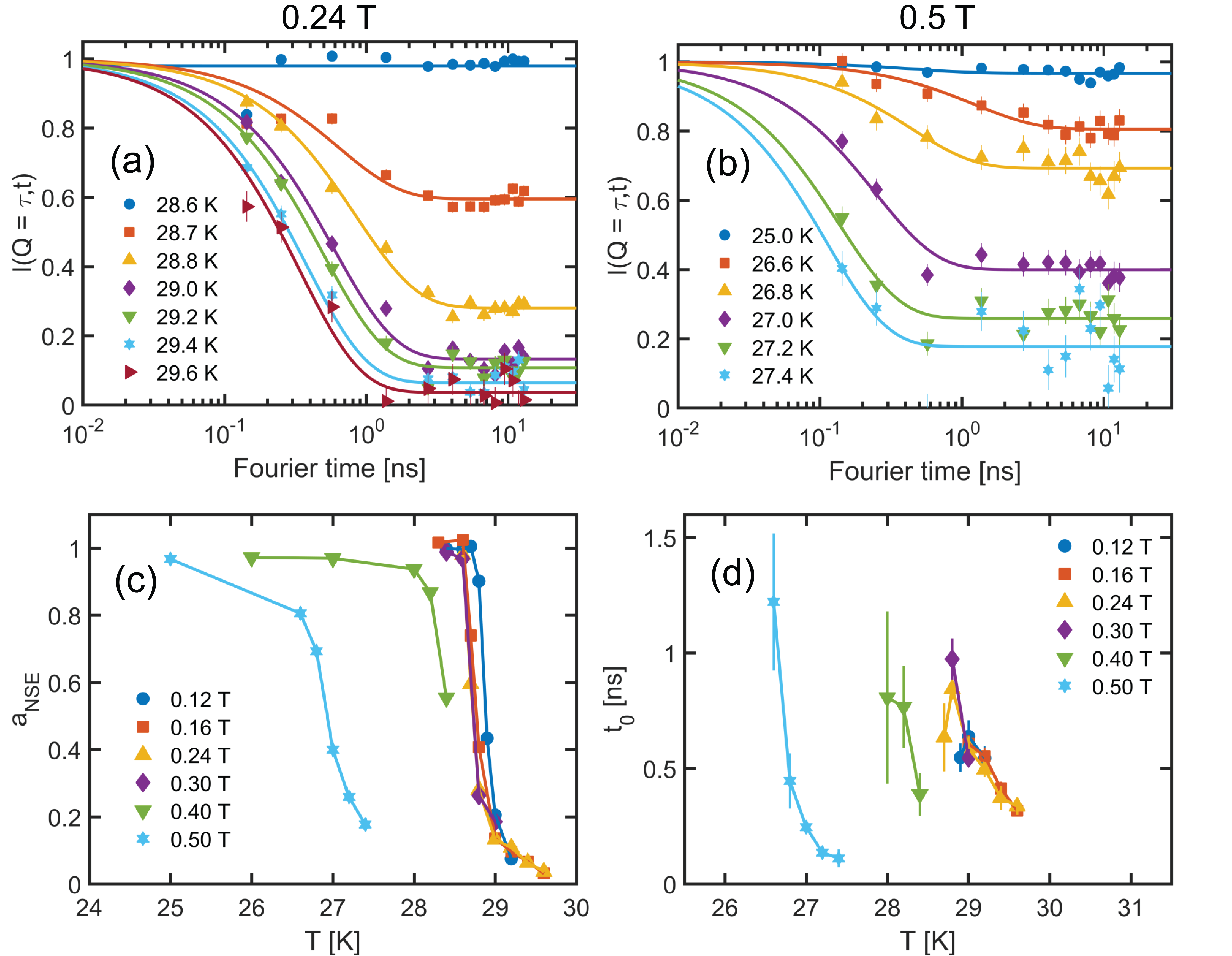}
\caption{Neutron Spin Echo spectroscopy results. Panel (a) and (b) show the intermediate scattering function $I(Q,t)$ measured at (a) 0.24~T and (b) 0.5~T.  Panel (c) shows the deduced elastic fractions $a_{\text {\tiny \it NSE}}$ and Panel (d) the relaxation times $t_0$ as a function of temperature for the magnetic fields indicated.}
\label{NSE}
\end{center}
\end{figure*}

We now switch to the NSE results that probed  the transition to the conical phase at twelve magnetic fields (0.12 $\leq B \leq $~0.5~T) applied perpendicular to $\vec{k_i}$, a configuration where the chiral correlations do not depolarize the scattered neutron beam \cite{Blume,Maleyev}. We note that for   $\vec{B} || \vec{k}_i$ the scattered beam was completely depolarized up to the highest magnetic fields indicating  that the probed correlations remain completely chiral and with the same topology as at zero field \cite{pappas2009,pappas2011}. 

Figs. \ref{NSE}(a) and (b) display NSE spectra for $B$ = 0.24~T and $B$ = 0.5~T respectively. The intermediate scattering function decays exponentially with a characteristic relaxation time $t_0$ to an elastic fraction $a_{\text {\tiny \it NSE}}$ that originates from the elastic contribution of the emerging Bragg peak:
\begin{equation}
I(Q,t)=(1-a_{\text {\tiny \it NSE}})\exp\left(-{t}/{t_0}\right)+a_{\text {\tiny \it NSE}}.
\label{eq:SE}
\end{equation}

\noindent The deduced values for $a_{\text {\tiny \it NSE}}$ and $t_0$ are displayed in Fig. \ref{NSE}(c) and (d). For all magnetic fields up to 0.4~T, $a_{\text {\tiny \it NSE}}$ evolves from $\sim$30\% to 100\% within 0.2~K following the abrupt appearance of the conical Bragg peaks seen in the increase of the scattered intensity with deceasing temperature displayed in Fig. \ref{SANS}(h). This is quantitatively similar to the zero field behavior \citep{pappas2009,pappas2011}. A  change sets-in at 0.5~T where the elastic fraction increases much slower from $\sim$30\% to 100\% within about 0.8~K.

A complementary picture is found for the relaxation times depicted in Fig. \ref{NSE}(d). For magnetic fields lower than 0.4~T, $t_0$ increases from $\sim$0.4~ns at $T_C$+1.5~K to $\sim$1~ns close to $T_C$, and this is very similar to the behavior observed at zero field  \citep{pappas2009,pappas2011}. At $B$ = 0.5~T, on the other hand, where the elastic fraction increase is slower, the temperature dependence of $t_0$ can be followed over a larger temperature range and is more pronounced than at low fields: $t_0$ is as low as 0.1~ns at 27.4~K, where the elastic fraction is already $\sim$20\%, and increases within 0.8~K by more than an order of magnitude to $\sim$ 1.2~ns at $T$ = 26.6~K. 


Both SANS and NSE show that the zero field sharp first order transition persists up to $B\sim $ 0.4~T. At higher magnetic fields the transition becomes gradual but this does not necessarily imply that it crosses over to a second order transition. This change in behavior occurs indeed when the transition line in the $B$ - $T$ phase diagram is no longer vertical \cite{bauer2012, muhlbauer2009} and might be at least partly attributed  to the demagnetization field, which would  spread the internal magnetic fields and thus smear the transition. We note that this should not affect the transition at lower fields where the transition line in the $B$ - $T$ phase diagram is almost vertical.

The precursor fluctuating correlations above $T_C$ remain chiral also under magnetic fields. However, for $B\gtrsim 0.2$~T they are gradually suppressed and the SANS intensity concentrates along the magnetic field direction. Consequently, the strength and degeneracy in space of the chiral fluctuations are affected by fields almost a factor of two weaker than those required to modify the first order nature of the transition. It thus appears that the precursor fluctuations and the first order nature of the transition are separate phenomena and are not accounted for by the Brazovskii approach \cite{janoschek2013}. Furthermore, the neutron scattering results presented above  do not confirm the existence of a tricritical point separating a first order transition at low fields from a second order one for fields exceeding 0.35~T. The existence of this point has been reported in the literature by magnetic susceptibility \cite{zhang2015} and specific heat \cite{bauer2013}. The discrepancy may arise from the  fact that our neutron scattering results probe selectively the helical correlations around $Q=\tau$ whereas magnetic susceptibility and specific heat probe the macroscopic behavior at $Q=0$. Indeed, high magnetic fields modify the balance between ferromagnetic  and helimagnetic correlations. This is illustrated by Fig. \ref{SANS}, which shows that at 0.35~T the helimagnetic correlations concentrate along the direction of the magnetic field and  are  at the same time the intensity considerably weakenes. This implies that the magnetic field induces ferromagnetic correlations at $Q$ = 0, thus outside our detection window, which may have significant impact on the $Q=0$ macroscopic behaviour. This effect starts at $B \sim 0.2$~T, the same field where changes start to be seen in the specific heat \cite{bauer2013}, and is amplified at higher magnetic fields. We speculate that it may be at the origin of a cross-over seen on magnetic susceptibility and specific heat, which has been interpreted as the signature of a tricritical point-like behaviour. 

To conclude, neutron scattering does not provide any indication for the existence of a tricritical point separating a first order transition, at low magnetic fields, from a second order one at high magnetic fields.  The transition remains of first order under magnetic field and is disconnected from the precursor fluctuating correlations above $T_C$.  These phenomena, which are both of general relevance to chiral magnetism, are thus not accounted for by the Brazovskii approach. Their understanding remains an open question calling for novel theoretical approaches in the future. \\

\acknowledgments
The authors thank P. Courtois for testing the MnSi single crystal with $\gamma$-ray Laue diffraction and the ISIS and ILL technical support staff for their assistance. They also acknowledge fruitful discussions with U. K. R\"{o}ssler, M. Mostovoy, G.R. Blake and T. Palstra. CP and LB acknowledges financial support by NWO Groot project LARMOR 721.012.102 and FQ from the China Scholarship Council. DS and TL acknowledge the support of the the US Department of Energy (DOE), Office of Science, Basic Energy Sciences, Materials Science and Engineering Division under Contract No. DE-AC02-07CH11358.

\bibliography{FeCoSi_Tc_Paper}		

\end{document}